\documentclass[Physsubmission, Phys]{SciPost}

\hypersetup{
    colorlinks,
    linkcolor={red!50!black},
    citecolor={blue!50!black},
    urlcolor={blue!80!black}
}

\usepackage[bitstream-charter]{mathdesign}
\DeclareSymbolFont{usualmathcal}{OMS}{cmsy}{m}{n}
\DeclareSymbolFontAlphabet{\mathcal}{usualmathcal}

\begin{document}

\begin{center}{\Large \textbf{
All About the Neutron from Lattice QCD\\
}}\end{center}

\begin{center}
Rajan Gupta\textsuperscript{1$\star$}
\end{center}

\begin{center}
{\bf 1} Los Alamos National Laboratory, Theoretical Division T-2, Los Alamos, New Mexico 87545, USA
\\
* CorrespondingAuthor rajan@lanl.gov
\end{center}

\begin{center}
\today
\end{center}


\definecolor{palegray}{gray}{0.95}
\begin{center}
\colorbox{palegray}{
  \begin{tabular}{rr}
  \begin{minipage}{0.1\textwidth}
    \includegraphics[width=23mm]{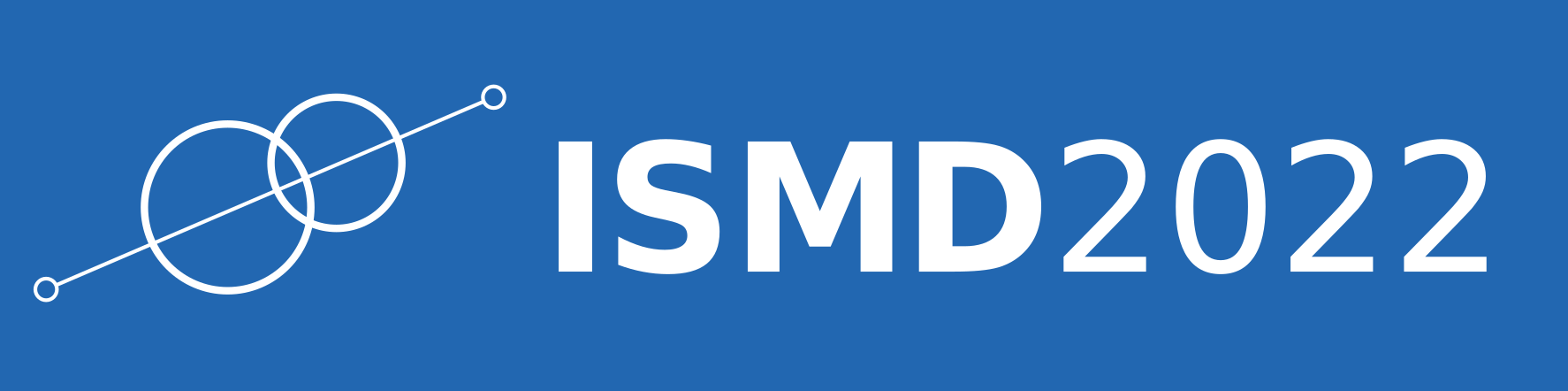}
  \end{minipage}
  &
  \begin{minipage}{0.8\textwidth}
    \begin{center}
    {\it 51st International Symposium on Multiparticle Dynamics (ISMD2022)}\\ 
    {\it Pitlochry, Scottish Highlands, 1-5 August 2022} \\
    \doi{10.21468/SciPostPhysProc.?}\\
    \end{center}
  \end{minipage}
\end{tabular}
}
\end{center}

\section*{Abstract}
{\bf
I describe how simulations of lattice QCD using the path integral
formulation provide the two basic quantum mechanical properties of
QCD, its ground state in which correlation functions are calculated,
and Fock state wavefunctions between which matrix elements of
operators are calculated.  Both constructs are stochastic, so
unfortunately one gets no intuitive picture or even qualitative
understanding of what they look like, nevertheless they contain and
display all the subtleties of the quantum field theory.  Today, these
simulations provide many quantities that are impacting phenomenology
and experiments. I illustrate the methods and the steps in the
analysis using, as examples, three observables: the isovector
charges of the nucleon, the contribution of the quark's intrinsic spin
to the nucleon spin, and the pion-nucleon sigma term.  }

\vspace{10pt}
\noindent\rule{\textwidth}{1pt}
\tableofcontents\thispagestyle{fancy}
\noindent\rule{\textwidth}{1pt}
\vspace{10pt}

\section{Lattice QCD}
\label{sec:LQCD}

\begin{figure}[hb]   
    \includegraphics[width=0.44\linewidth]{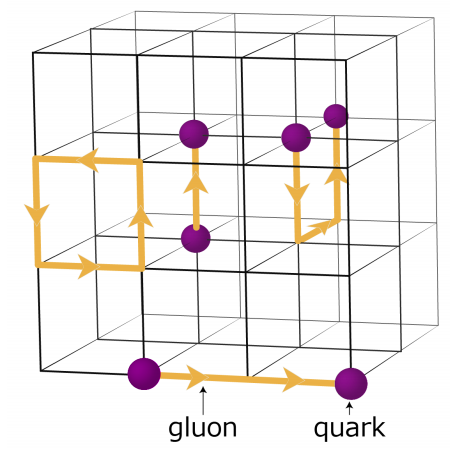} 
    \includegraphics[width=0.44\linewidth]{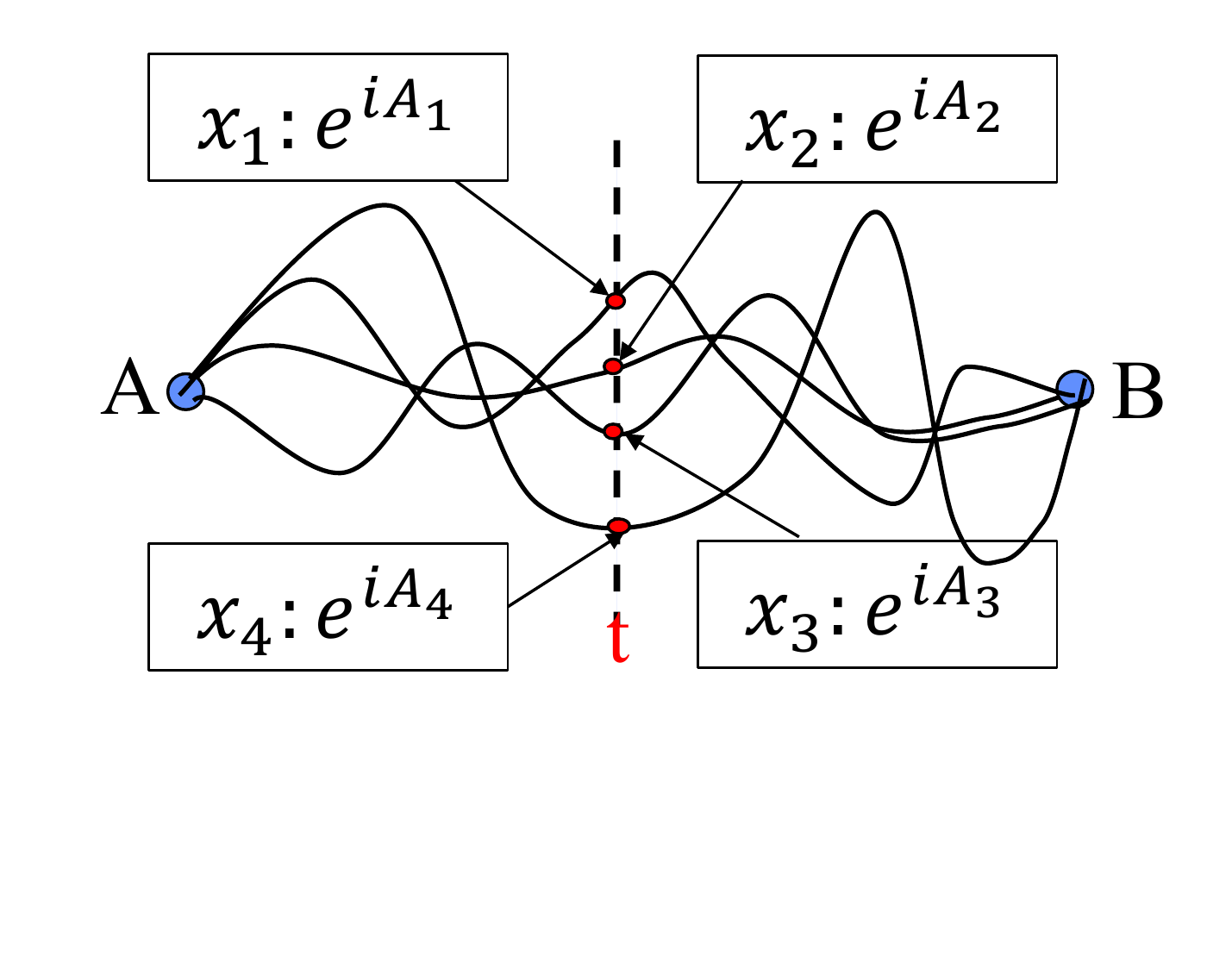} 
\vspace{-0.1in}
\caption{(Left) Discretization of QCD on a hypercubic lattice with
  quark fields placed on sites and the gluon fields $A_\mu(x)$ on
  directed gauge links via SU3 matrices ${\cal U}_{x,\mu} = e^{iag
    A_\mu^a \lambda^a}$ where $g$ is the gauge coupling and
  $\lambda^a$ are the Gell-Mann matrices. This lattice formulation preserves
  gauge invariance of the continuum theory. (Right) Illustration
  of the path intergral formulation of quantum mechanics of a particle
  moving between points A and B. Each path $i$ has coordinate $x_i$ at
  time $t$ and is weighted by $e^{iA_i}$, where $A_i$ is the action. 
  All possible paths connecting A and B contribute.\looseness-1}
\label{fig:lattice}
\end{figure}

The field of lattice QCD (LQCD)---the theoretical rigorous path
integral formulation of QCD discretized on a 4D hypercubic grid by
Wilson, and used to provide non-perturbative
predictions~\cite{Wilson:1974sk}---has come of age. This formalism
converts quantum field theories into statistical mechanics systems,
for example, the 3+1 dimensional QCD in Minkowski time becomes, after a Wick rotation to 
Euclidean time, a classical system of gluon and quark fields on a 3+1 dimensional
lattice in Euclidean time (see Fig.~\ref{fig:lattice}).  Numerical
simulations of it~\cite{Creutz:1984mg} are providing first principle
results with control over all systematic uncertainties for a large
number of physical observables that elucidate the standard model and
probe physics beyond it.  The Flavor Lattice Averaging Group
(FLAG~\cite{FLAG}) provides a community based evaluation of quantities
that are considered
robust~\cite{Aoki:2019cca,Aoki:2021kgd}\footnote{The chapter on
  Nucleon Matrix Elements (NME) in these reports provides an introduction
  to the many issues relevant to the calculation of NME
  discussed in this talk and contains an extensive list of
  references for the interested reader. Very often I will just refer
  to the FLAG reports with the understanding that a list of pertinent
  references is already collected there.}.  With improvements in
numerical algorithms and increasing computing resources, the errors on
these quantities are being reduced steadily, and many more quantities
are being added to the list. At the same time, the need for new ideas
for a big leap forward is also evident. In this writeup, I will
present an idiosyncratic mixture of topics, starting with explaining
what simulations of LQCD give us, and then highlight successes and, at
the same time, the need for new ideas for sub-percent precision
predictions of the properties of nucleons.\looseness-1

I begin with a brief recap of the path integral formulation of a
particle moving in time. Figure~\ref{fig:lattice} (right) illustrates
some of the paths that contribute to the quantum mechanical amplitude
for the particle to go from point A to point B. In fact, all paths one
can draw between those two points contribute.  Each path $n$ has a
weight $e^{iA_n}$, where $A_n$ is the action of that path. The value of an
observable, say the position $x$ at a prescribed time $t$ is given by
the expectation value $\sum_n x_n e^{iA_n}$/$\sum_i e^{iA_n}$ where
$x_n$ is the position at time $t$. The
amplitudes interfere and, typically, the path with the smallest action
gives the largest contribution. In the classical limit, this path
converges to that predicted by Newton's equations.  
Using this example, I now motivate how simulations of lattice QCD give us
the analogues of the ``paths'', the correlation functions
corresponding to observables such as $x$, and the wavefunctions within
which matrix elements (ME) of operators ($\to$observables) can be
calculated.

In simulations of lattice QCD (LQCD), and of gauge field theories in
Euclidean time in general, the analogue of the paths are gauge
configurations. Each LQCD configuration ${\cal C}_i$ is a
specification of the 12 independent entries in each SU(3) matrix,
${\cal U}_{x,\mu}$, assigned to each link of the lattice (see
Fig.~\ref{fig:lattice}). The ${\cal C}_i$ have a weight $e^{-A_i}$,
where $A_i$ is the Euclidean QCD action calculated on 
configuration ${\cal C}_i$. It is a functional of all the ${\cal U}_{x,\mu}$. Note,
the quark fields are formally integrated out as discussed below,
leaving only gauge fields as dynamical variables. Configurations are
generated using Markov Chain Monte Carlo Methods with importance
sampling and the Metropolis accept/reject step~\cite{Creutz:1984mg}.
Conceptually, this algorithm for the generation of the ${\cal C}_i$ is
the same as used in the classical simulations of spin models, however,
simulations are computationally expensive because on a $100^4$
lattice, there are $4 \times 12 \times 10^8$ independent variables
that specify a ${\cal C}_i$ (entries in all the SU(3) link matrices)
and evaluating the action $A_i$ is expensive. The full set of the ${\cal
  C}_i$ (called an ensemble) and their associated $A_i$ provides us
the fully quantum mechanical ground state of the field theory, 
{\it albeit} stochastically since only a finite ${\cal C}_i$ are sampled
in practice. The action $A$ is characterized by the input parameters
of the simulations: quark masses $m_i$ with $i \in \{u, d, s, c\}$
flavors, lattice spacing $a$ (equivalently the gauge coupling $\beta = 6/g^2$ via
dimensional transmutation)~\cite{Creutz:1984mg}, and the lattice
volume $L^3 \times T$.

The full set of configurations is the same independent of
the action, $A(m_i, a)$, i.e., of all input parameters $\{m_i, a\}$, 
and depends only on the number of gauge links ($4\times L^3 \times T $) 
and the values they can take. There are $\infty^2$ of them: infinite
number of variables in the limit the volume $L^3 \times T \to \infty$,
and each variable is continuous valued between $\{-1,1\}$. If
even a significant subset of these were needed to calculate
observables, precision would not be achieved.\looseness-1

\section{Correlation Functions and Observables}
\label{sec:CF}

Expectation values of observables $O$ are obtained from ensemble
averages, $\sum_i \Gamma_\alpha e^{-A_i}$/$\sum_i e^{-A_i}$, of
correlation functions $\Gamma_\alpha$ measured on the ${\cal C}_i$. What saves
us from having to consider the $\infty^2$ configurations is that
the weight $e^{-A_i}$ is so very highly peaked about the minimum of
$A$ that $10^3 -10^7$ (depending on the observable) importance sampled
and statistically independent configurations suffice to yield
expectation values with sufficient precision.  The location of the
peak of the distribution (minimum of $A$) changes with the input
parameters, i.e., with the $\{m_i, a, L,T\}$.

In practice, data (expectation values of correlation functions) are
obtained on many ensembles with different $\{m_i, a, L\}$ so that the
limits $a\to 0$, $m_i $ to their physical values set by the
experimental values of the masses $M_\pi$, $M_N$, $M_\Omega$ and
$M_D$, and $L \to \infty$ can be taken to obtain physical results.
Typical in current simulations, $m_s$ and $m_c$, being sufficiently
heavy, are already tuned to their physical values before starting
production, and only $m_{u,d}$ are varied.  In the isospin symmetric
limit, $m_{u}=m_{d}$, it is typical to represent the common light
quark mass ${\overline m}_{ud}$ by the corresponding value of $M_\pi$.
Again, ${\overline m}_{ud}$, is tuned before starting production
runs. Today, we can perform simulations at $M_\pi = 135$~MeV, but very
often data are also obtained at a number of heavier values of
${\overline m}_{ud}$ (equivalently $M_\pi$), and then extrapolated to
$M_\pi=$ 135 MeV using ansatz motived by chiral perturbation
theory. These ideas will be illustrated by the calculations/results
reviewed later.

An essential simplification, in fact one that allows simulations of
QCD on classical computers in the first place, is that the fermion
action for each flavor $q$, 
\begin{equation} 
A_F = {\bar \psi}D \psi = {\bar \psi} (\gamma_\mu (\partial_\mu + i g A_\mu) + m_q) \psi \to 
{\bar \psi}(x+a\hat \mu) (\gamma_\mu U_{x,\mu}) \psi(x) + m_q {\bar \psi}(x)\psi(x)  \,,
\end{equation}
is bilinear in the quark fields. Here $D$ is the Dirac operator, which
on the lattice is a $(3\times 4 \times L^3 \times T)^2$ complex matrix
that depends only on the $U_{x,\mu}$.  The fermions can, therefore, be
integrated out exactly from the path integral but contribute the
determinant of the Dirac matrix, ${\rm Det}{{\cal D}_f}$, for each
flavor $f$ to the Boltzmann weight, which becomes $\prod_f ({\rm Det}
{{\cal D}_f)} \, e^{-A_{G}} = e^{-A_{G} + \sum_f{\rm Ln} {\rm Det}
  {\cal D}_f} $ where $A_G$ is the gauge action. They, therefore,
continue to impact the $A(m_i, a, L)$, and thus the position of the
peak of the distribution specifying the ground state and fluctuations
about it.  Calculating the contribution of the determinant to the
Boltzmann factor used to generate the configurations makes simulations
expensive but does not pose a formal obstruction.  The inverse of
$D_f$, a sparse matrix, is the all-to-all Feynman quark propagator.
One column (or row) of ${\cal D}_f^{-1} $ is the point-to-all $S_F$
used to construct correlations functions.

To construct the 2- and 3-point correlation functions of operators
composed of quark fields, even though we have integrated them out
formally, consider the time-ordered product $\cal T$ of the pion
interpolating operator ${\bar u} \gamma_5 d$ and the axial current ${\bar d} \gamma_\mu \gamma_5 u$:
\begin{equation}
\Gamma^2_\pi = \langle \ {\cal T} ({\bar d} \gamma_5 u|_\tau \  {\bar u} \gamma_5 d|_0 ) \  \rangle ; \quad 
\Gamma^3_\pi = \langle \ {\cal T} ({\bar d} \gamma_5 d|_\tau \  {\bar d} \gamma_\mu \gamma_5 u|_t \  {\bar u} \gamma_5 d|_0 ) \rangle \,,
\label{eq:TOP} 
\end{equation}
with the assumption that all three (actually only 2 since momentum is
conserved in LQCD) operators have been projected to zero momentum for
simplicity, thus leaving only the time index.  The notation $\langle
\ \cdots \ \rangle $ implies ensemble average over the ${{\cal C}_i}$.
At the same time as integrating out the quarks, one can perform a Wick
contraction of the fields in Eq.~\ref{eq:TOP} to get
\begin{equation}
\Gamma^2_\pi = \langle \ S_F(0,\tau) \gamma_5 S_F(\tau,0) \gamma_5 \  \rangle ; \quad 
\Gamma^3_\pi = \langle \ S_F(0,\tau) \gamma_5 S_F(\tau,t) \gamma_\mu \gamma_5 S_F(t,0) \gamma_5  \rangle
\end{equation}
where $S_F$ (a column of the inverse of the Dirac matrix ${\cal
  D}$ calculated using iterative Krylov solvers) is the Feynman
propagator from a point source to all lattice points.  Now using the
hermiticity property of the Dirac action and its inverse, $S_F(0,\tau)
= \gamma_5 S_F^\dagger (\tau,0) \gamma_5 $, we get
\begin{equation}
\Gamma^2_\pi = \langle S_F(0,\tau) S_F^\dagger (0,\tau)   \rangle ; \quad
\Gamma^3_\pi = \langle S_F(0,\tau) S_F^\dag (\tau,t) \gamma_\mu \gamma_5 S_F^\dag (0,t)   \rangle
\label{eq:2and3pion}
\end{equation}

\begin{figure}[h]  
    \includegraphics[width=0.32\linewidth]{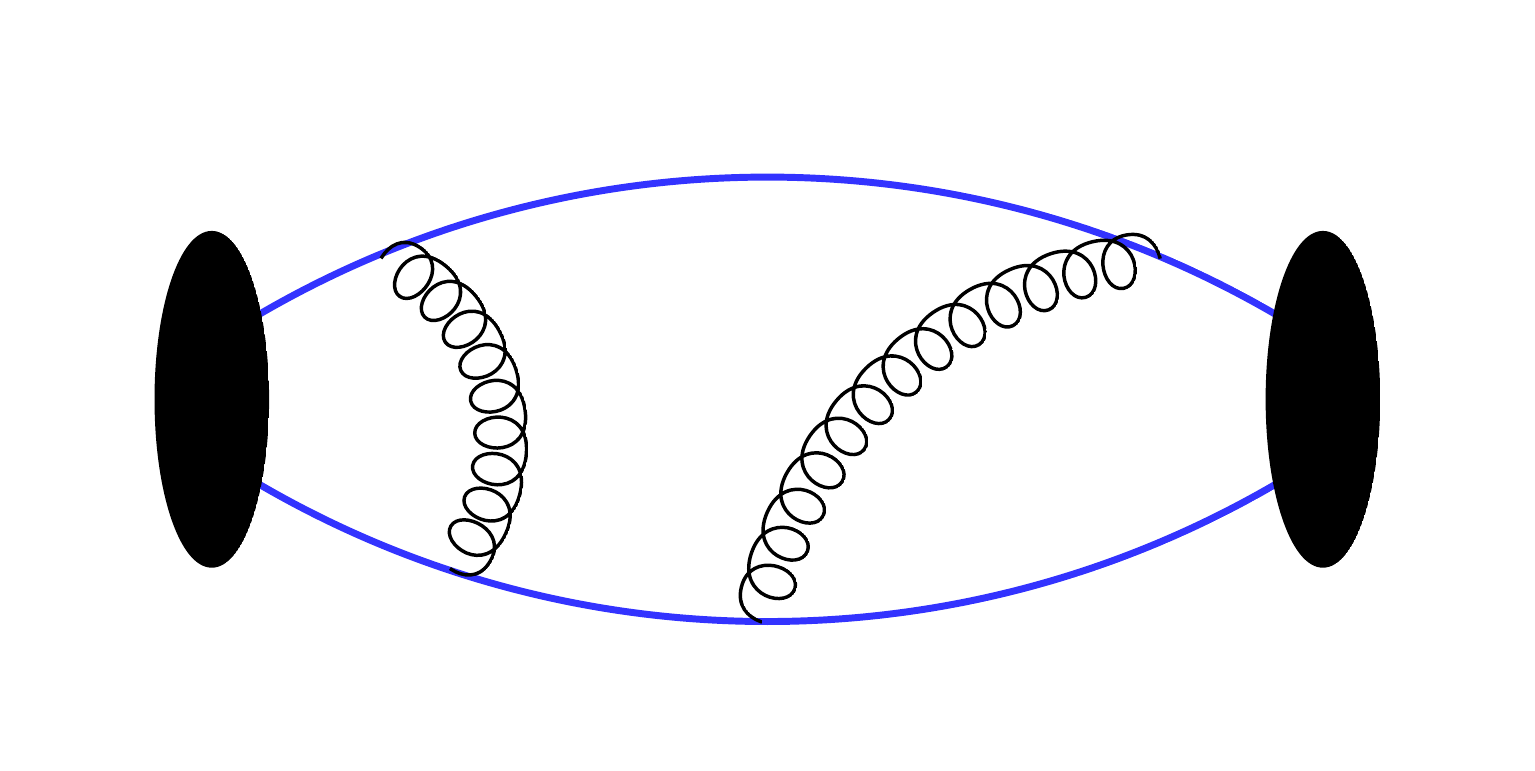}
    \includegraphics[width=0.32\linewidth]{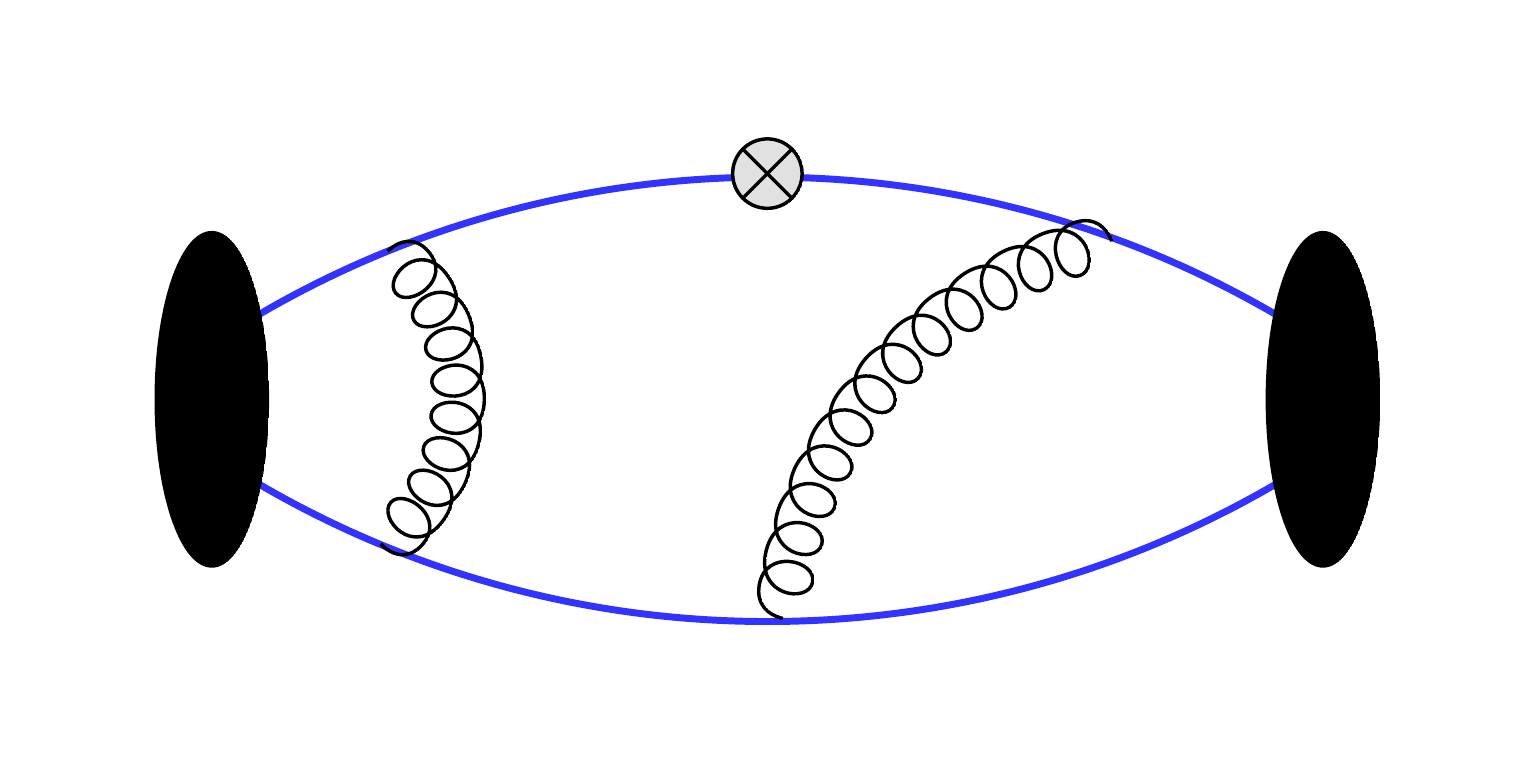}
    \includegraphics[width=0.34\linewidth]{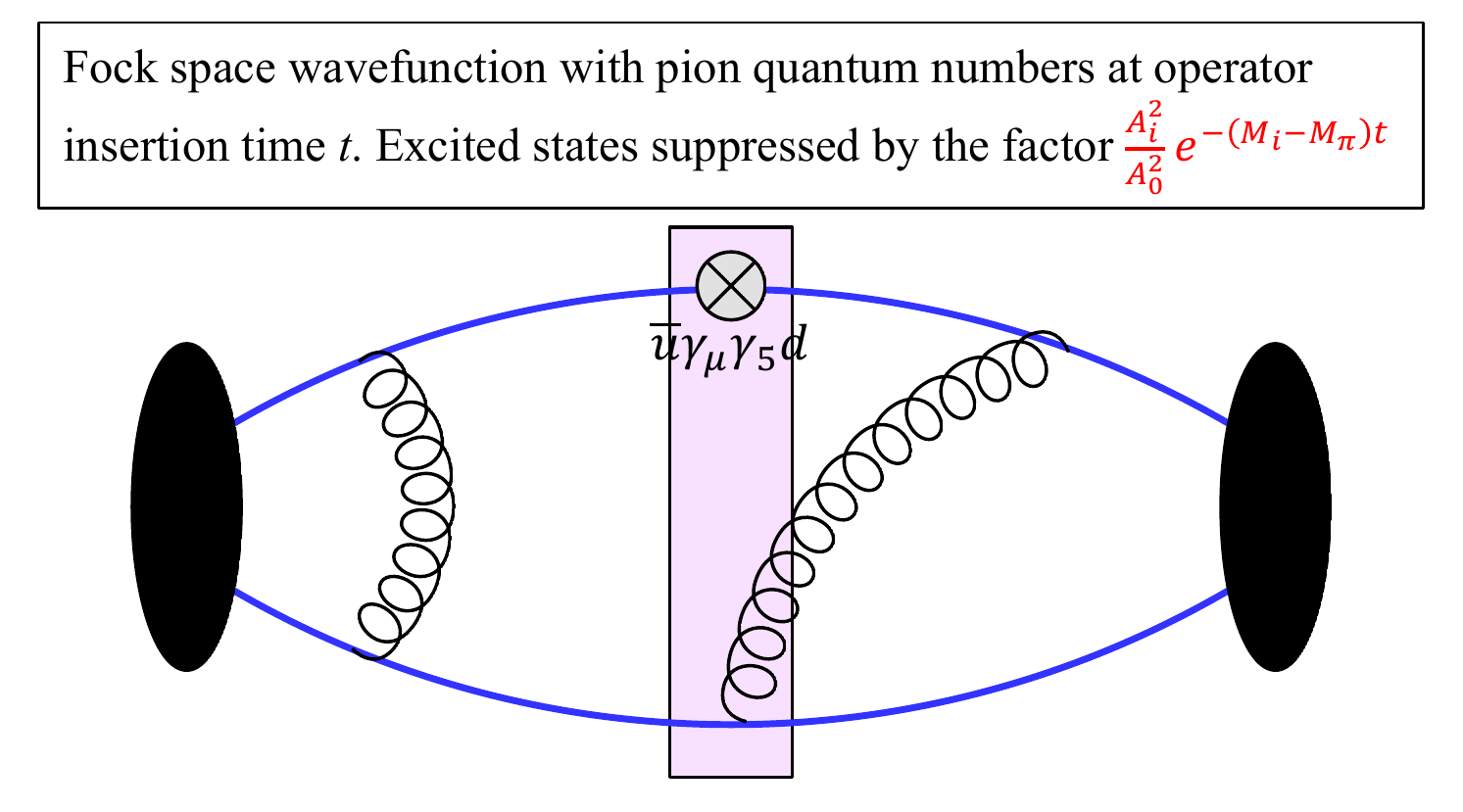}
\vspace{-0.1in}
\caption{Illustration of quark-line diagrams for 2-point (left) and
  3-point functions for the pion (middle and right).  The gluon lines are just
  for illustration and to remind the reader that all orders of gluon
  exchanges are implicit in these diagrams. (Right) The axial current,
  shown by $\bigotimes$, is inserted at intermediate Euclidean time
  $t$ and with momentum $\vec q$. The ensemble average in LQCD
  simulations creates the stochastic Fock state wavefunction at each time $t$, indicated
  by the pink band, and the operator 
  causes transitions between the various ``pion'' states of the
  transfer matrix. The extraction of these matrix elements (and thereby
  pion's axial form factors) are obtained from fits to these 2- and
  3-point correlation functions using
  Eq.~\eqref{eq:SF2and3pion}. \looseness-1}
\label{fig:2and3ptPi}
\end{figure}

Thus performing the Wick contraction replaces the quark fields in
correlation functions in terms of $S_F = {\cal D}^{-1}$, which depends
only on the gauge links. {\it In short, both in the generation of the
  configurations and in the calculation of correlation functions, the
  quark fields are integrated out exactly. } The expressions in
Eq.~\ref{eq:2and3pion} for the pion correspond to the quark line
diagrams shown in Fig.~\ref{fig:2and3ptPi}, whose expectation values
give the desired non-perturbative correlation functions.

The question I hope you are dying to ask is how does the quark
propagator, $S_F$, calculated on a given configuration and combined to
form the quark-line diagrams shown in Fig.~\ref{fig:2and3ptPi}, know
anything about the non-perturbative propagation of the pion or any of
the thousands of possible states of QCD, the analogous quark-line
diagrams for which are obtained by simply changing the interpolating
operators.  As already explained, the interpolating operators create
states with given quantum numbers, for example ${\bar d} \gamma_5 u$
or ${\bar d} \gamma_0 \gamma_5 u$ for the pion. These are propagated in
time by the transfer matrix.  The non-perturbative properties of the
propagating pion and its dynamics arise from the coherent addition of
those gauge fluctuations on each configuration that correspond to a
pion propagating. The miracle of the ensemble average is that only fluctuations
corresponding to states with pion quantum numbers survive. In short,
the frothing vacuum has all possible fluctuations present, and the
ensemble average picks up those that conform to the quantum numbers of
the state created by a given interpolating operator.

Another interesting aside is that $\Gamma^2_\pi = \langle | S_F(0,\tau) |^2
\rangle $ in Eq.~\ref{eq:2and3pion} is a positive definite quantity.  So you may ask--how
can averaging over only a small ``working ensemble'' give a precise
unbiased result? The answer lies in the fact that configurations importance sampled 
according to the Boltzmann weight $e^{-A}$ provide an unbiased 
approximating to the full ensemble (path integral). Clearly, to improve statistical
precision, one needs to enlarge the ``working ensemble''.

Now we come to the last part of the introduction to LQCD--how does one
get physics from correlation functions such as those in
Eq.~\ref{eq:2and3pion}? For this we invoke the spectral decomposition
of $\Gamma^2_\pi$ and $\Gamma^3_\pi$, i.e., the insertion of a
complete set of ``pion'' states $|\pi_i\rangle$ at each intermediate time step, and
the evolution between time steps given by the transfer matrix. The
result is
\begin{equation}
\Gamma^2_\pi = \sum_i |\langle 0 |{\hat \pi_i} | \pi \rangle |^2 \ e^{-E_i \tau} ; \quad
\Gamma^3_\pi = \sum_{i,j} \langle 0 |{\hat \pi} | \pi_i \rangle^\ast \ e^{-E_i (\tau-t) }
\langle \pi_i |{\hat A_\mu} | \pi_j \rangle     e^{-E_i (t-0)}   \langle \pi_j |{\hat \pi} | 0 \rangle \,, 
\label{eq:SF2and3pion}
\end{equation}
where ${\hat \pi}$ is the pion interpolating operator, and the sum over $\{i,j\}$ is 
over all the states of the Transfer matrix with the quantum numbers of the pion. Such  
decompositions of $\Gamma^n$  hold for all interpolating operators and 
the states they couple to. Simply replace the symbol $\pi$
by the state of interest. By fitting $\Gamma^2_\pi $ versus $\tau$, we
can extract the amplitudes, $|\langle 0 |{\hat \pi} | \pi_i \rangle |^2
$ and the energies $E_i$ for all the "pion" states that couple to ${\hat
  \pi}$. In the limit $\tau \to \infty$, only the ground (lowest) state
contributes, and for ${\hat \pi} = {\bar d} \gamma_4 \gamma_5 u$, one
gets from $\Gamma^2$ the pion decay constant $F_\pi$ since $ |\langle
0 |{\hat \pi} | \pi_0 \rangle |^2 = M_\pi^2 F_\pi^2$, and its mass $M_\pi$.
Thus 2-point functions give us the amplitudes for creating the state and the spectrum of the
theory (actually, in discrete time, of the Transfer Matrix).\looseness-1

Next, consider $\Gamma^3_\pi$.  It has an additional operator,
${\hat A_\mu} $, sandwiched between the pion creation and 
annihilation operators. The propagating tower of "pion" states, $| \pi_i \rangle $,  interact with 
current ${\hat A_\mu} $ at time $t$, which causes transitions between these states with strength 
given by the matrix element, $\langle \pi_i |{\hat A_\mu} | \pi_j
\rangle $. These ME can be isolated from the fit to $\Gamma^3_\pi$ since all the
other terms, can, in principle, be determined from the fit to
$\Gamma^2_\pi$. It is easy to check that as $\tau \to \infty$, the ME 
within the ground state of the pion is given by the ratio $\Gamma_3 / \Gamma_2$.

If any two operators in Eq.~\ref{eq:2and3pion} are projected to $\vec
p=0$ (LQCD conserves momentum), then we get the axial charge of the
pion. If the axial current ${\hat A_\mu} $ inserts momentum $\vec q$
and one of the pion interpolating operator removes it, we get the
axial form factor describing the semileptonic decay of pions with
Euclidean momentum transfer $Q^2 = |\vec q |^2 - (E_f - E_i)^2$.

Once data for $\Gamma^2$ and $\Gamma^3$ are collected at a number of
values of $\{m_i, a, L\}$, and fit using Eq.~\ref{eq:SF2and3pion} to
get data for decay constants, energies $E_i$ and matrix elements,
their physical valus are obtained by a simultaneous extrapolation:
$M_\pi \to 135$~MeV, $a \to 0$, and $L\to \infty$ using physics
motivated ans\"atz. This extrapolation is common to all LQCD
calculations as illustrated below.

\section{Renormalization of Lattice Operators}
\label{sec:Renorm}

We can write down a number of equally good lattice operators $O^{{\rm
    latt},n}$ that should give the same results in the continuum
limit. At finite $a$, the results will differ due to their relative
normalization and different discretization errors, over and above the
known differences coming from the amplitudes, such as $\langle \pi
|{\hat \pi_i} | \Omega \rangle $ if different interpolating operators
${\hat \pi_i}$ are used.  Lattice renormalization factors, $Z^{\rm
  latt}_{O_n}$, relate the different $O_n$ at a given $a$, and their
scaling behavior as $a\to 0$.  Results with renormalized operators,
say $Z_{A_n}^{{\rm latt}} A^{{\rm latt}}_{\mu,n}$, should agree in the
continuum limit.\looseness-1

The experimental results presented by phenomenologists typically use a
scheme such as $\overline {MS}$ and a convenient scale such as $2$~GeV
above which perturbation theory is considered reliable. To translate
the lattice result to the $\overline {MS}$ scheme at, say, 2 GeV is a two
step process. First one calculates the lattice factors $Z^i_{O}$ in
some scheme (currently two popular ones are the regularization
independent [symmetric] momentum schemes labeled RI-MOM and
RI-$s$MOM~\cite{Martinelli:1994ty,Sturm:2009kb}), and a second
calculation that relates them to $\overline {MS}$ that is typically
done in the continuum using perturbation theory, as is the factor for 
running in the continuum to a specified scale, say, 2 GeV. 

For the calculations described here, the renormalization factors are
well-determined.  Many other operators, such as the CP-violating
Weinberg and quark chromo EDM operators of dimension 6 and 5,
respectively, have divergent mixing with lower dimension operators.
Cnstructing finite renormalized versions to use in simulations is very
non-trivial~\cite{Bhattacharya:2015rsa}. In fact, for these two
operators, it is still an open problem.


\section{Nucleon Correlation Functions}
\label{sec:LQCD_nuc}

\begin{figure}[h]  
    \includegraphics[width=0.32\linewidth]{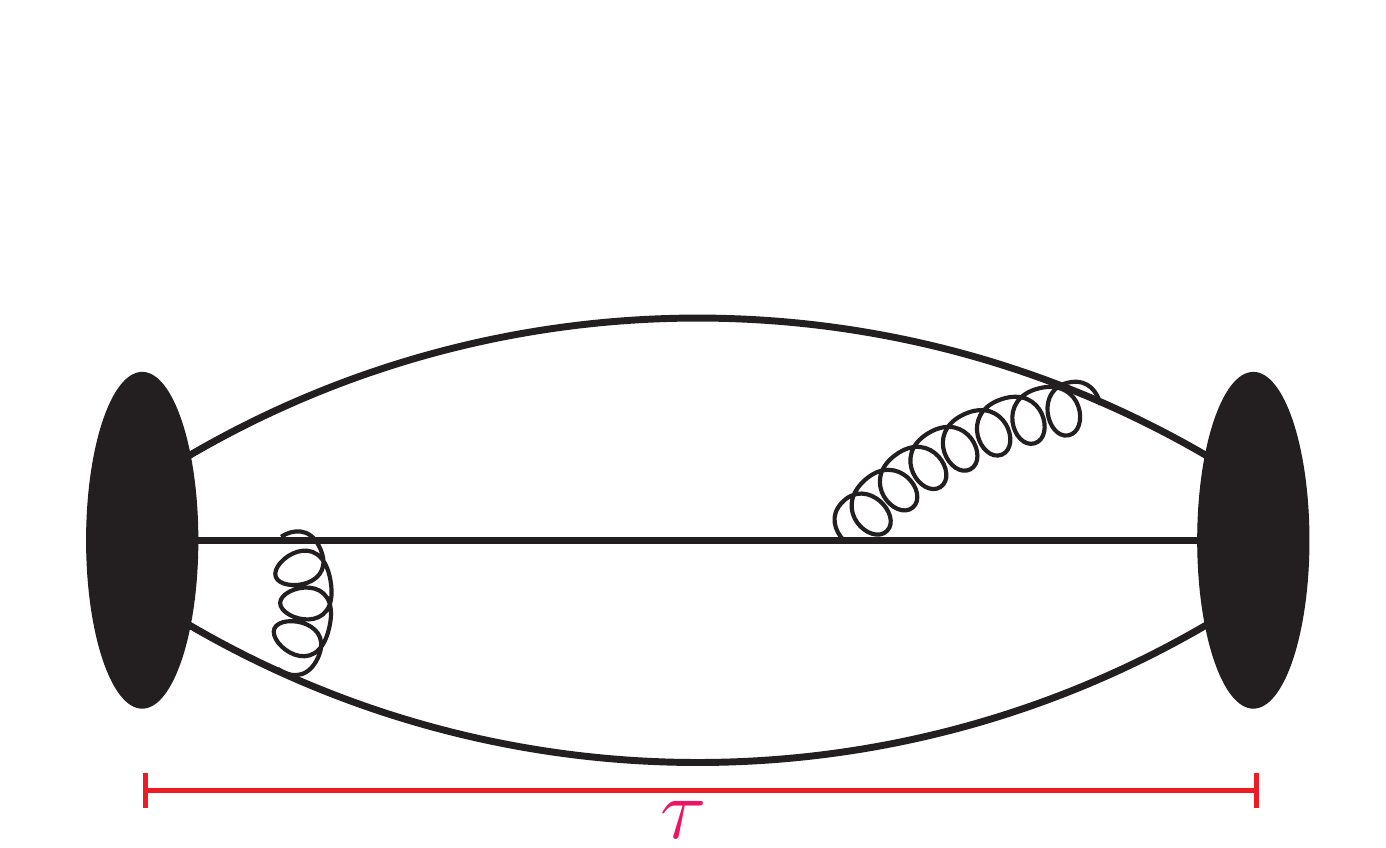}
    \includegraphics[width=0.32\linewidth]{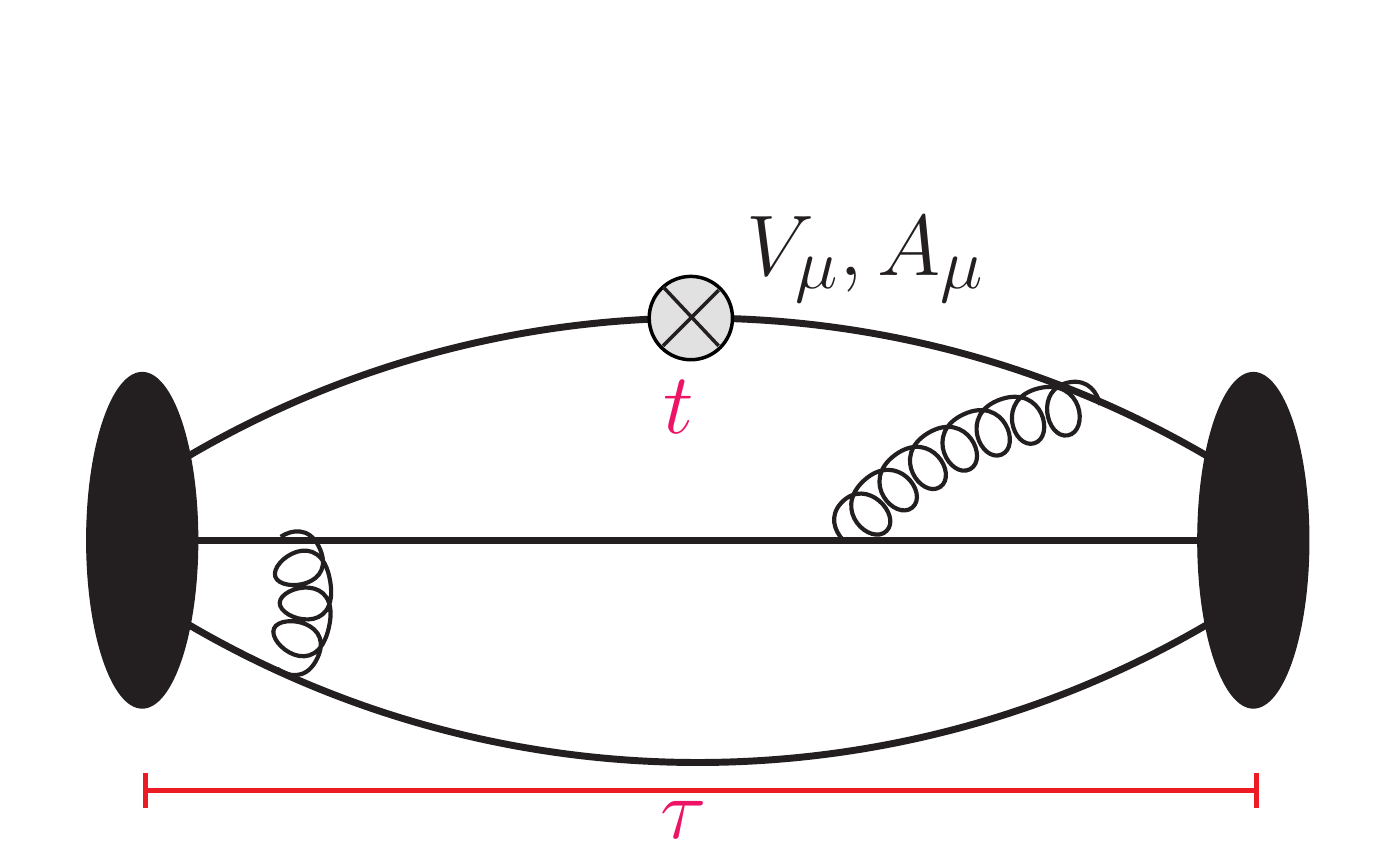}
    \includegraphics[width=0.32\linewidth]{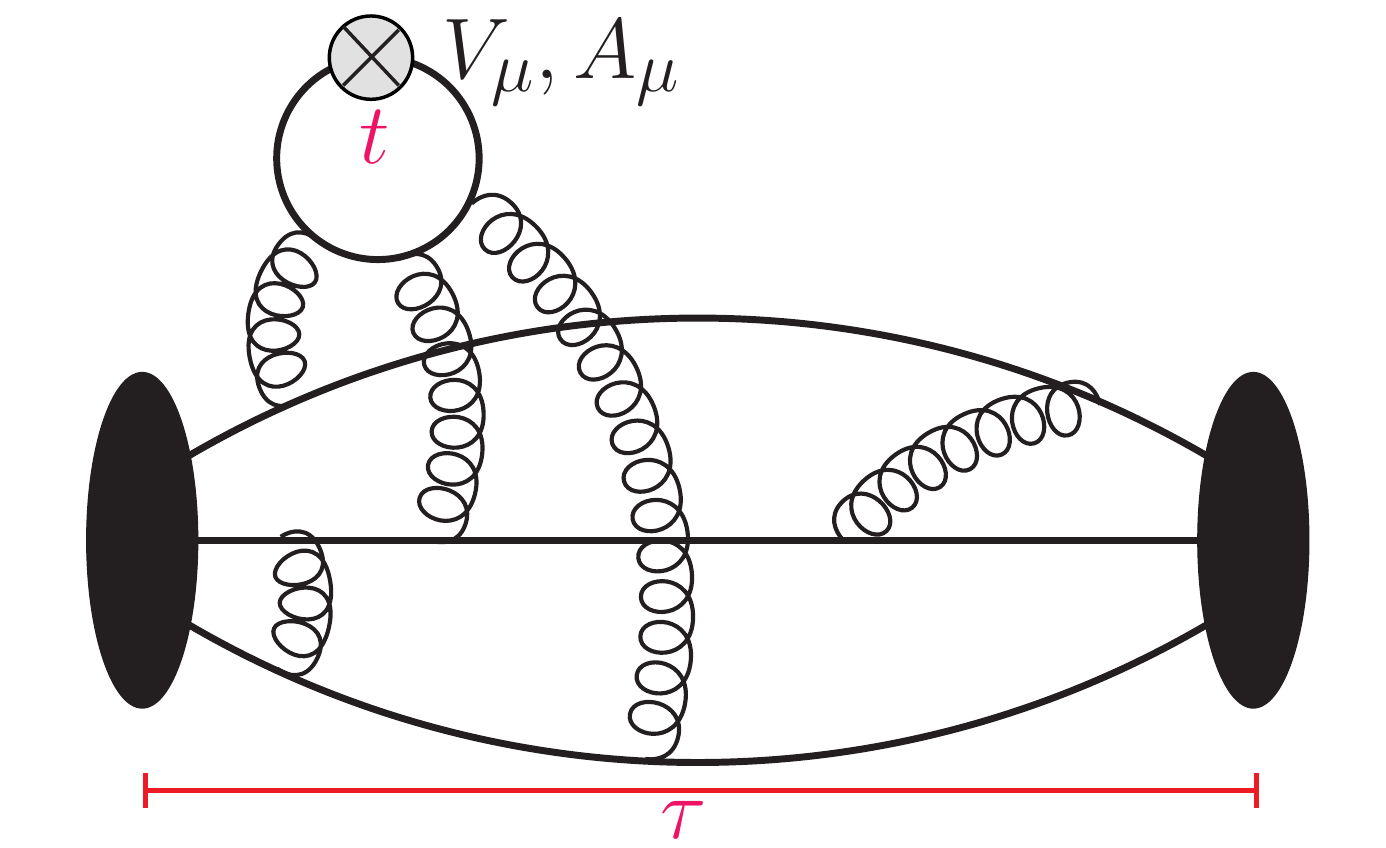}
\vspace{-0.1in}
\caption{Illustration of quark-line diagrams for nucleon 2-point function, $\Gamma^2$, (left);
  connected 3-point $\Gamma^3$ with insertion of iso-vector vector (axial) operator $\bar u \gamma_\mu
  d$ ($\bar u \gamma_5 \gamma_\mu d$) at intermediate Euclidean time
  $t$ (middle); and the additional disconnected
  contribution for flavor diagonal vector (axial)
  operators $\bar q \gamma_\mu q$ ($\bar q \gamma_\mu \gamma_5 q$)
  (right). The vector/axial form factors of the nucleon are obtained from the ground state ME 
  $\langle N_0 | {\hat {V}_\mu},{\hat A_\mu} | N_0 \rangle $ extracted from 
  $\Gamma^3$ with ${\hat {V}_\mu},{\hat A_\mu}$ 
  inserted with momentum $\vec q$. \looseness-1}
\label{fig:2and3ptN}
\end{figure}

The quark line diagrams 
for the nucleon 2- and 3-point functions are shown in Fig.~\ref{fig:2and3ptN}. 
Formally, the mechanics of the lattice calculation is very similar to that 
for the pion, however there are two very important differences:
\begin{itemize}
\item The signal to noise ratio falls exponentially as $\sim e^{-(E_N
  - 1.5 M_\pi)\tau}$ in all nucleon correlation functions, whereas the
  pion has no degradation. Typical data show that for $\Gamma^2_N$ a
  good signal extends to about $2$~fm and for $\Gamma^3_N$ to about
  $1.5$~fm with $O(10^6)$ measurements~\cite{Park:2021ypf}.
\item For a number of matrix elements, excited states contributions
  (ESC) from towers of multi-hadron excited states, $N \pi$, $N\pi
  \pi$, $\ldots$ labeled by their relative momentum, are enhanced and
  still large at 1.5 fm~\cite{Park:2021ypf}.  Their energies begin at
  about 1200 MeV, much below radial excitations. Fully 
  removing these ESC in fits to $\Gamma^2$ and $\Gamma^3$ remains a challenge for many
  observables.\looseness-1
\end{itemize}

To determine various quantities, we use appropriate probes.
Changing the operator to a scalar, ${\hat S} = {\bar d} u$ or
tensor, ${\hat T} = {\bar d} \sigma_{\mu \nu} u$, gives us nucleon's scalar
and tensor charges that are also probed in precision measurements of
neutron decay distributions~\cite{Bhattacharya:2011qm}. One link operators give us the
momentum fraction, helicity and transversity moments~\cite{Mondal:2020ela}. And the
list continues.

Having laid out, hopefully, an intuitive introduction to simulations
of lattice QCD, I now discuss three calculations in order of
increasing complexity.

\section{Isovector Charges of the Nucleon}
\label{sec:isovectorcharges}

The iso-vector axial, scalar, tensor charges of the nucleon,
$g_A^{u-d}$, $g_S^{u-d}$, and $g_T^{u-d}$, probed in the $N \to P$ decay, 
are extracted from $\Gamma^3(\vec p = 0)$, i.e., from the forward matrix element 
\begin{align}
\langle P({\vec p}=0,s') |  Z_O \, \bar{u} X_O  d|_{\vec q = 0} | N ({\vec p}=0,s) \rangle =  
g_O   \ u_P (0,s') \,  X_O  \, \bar{u}_N (0,s)  \,.
\label{eq:gAdef}
\end{align}
with Dirac matrix $X_O = \gamma_\mu \gamma_5,\ \mathbb{1}, \ \sigma_{\mu \nu}$
specifying the insertion of the axial, scalar and tensor operators at
zero momentum transfer. For these iso-vector charges, only the connected quark line diagram (middle
panel in Fig~\ref{fig:2and3ptN}) contributes in the isospin
symmetric limit, and results for the proton and the neutron are the same. 

\begin{figure}[h]  
    \includegraphics[width=0.44\linewidth]{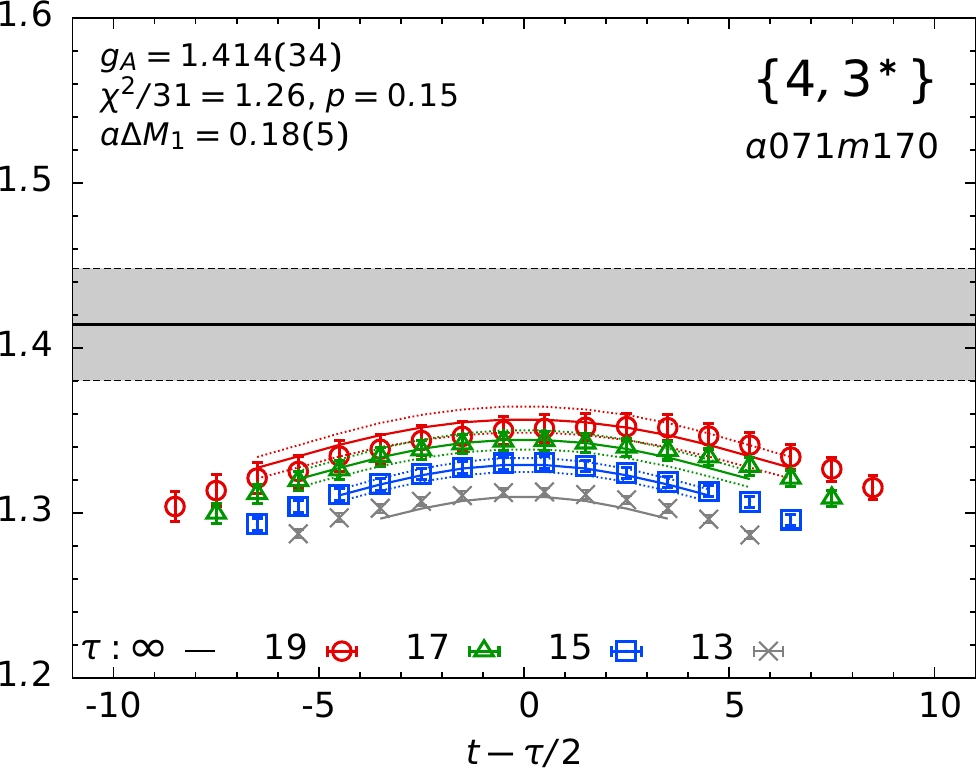} 
    \includegraphics[width=0.44\linewidth]{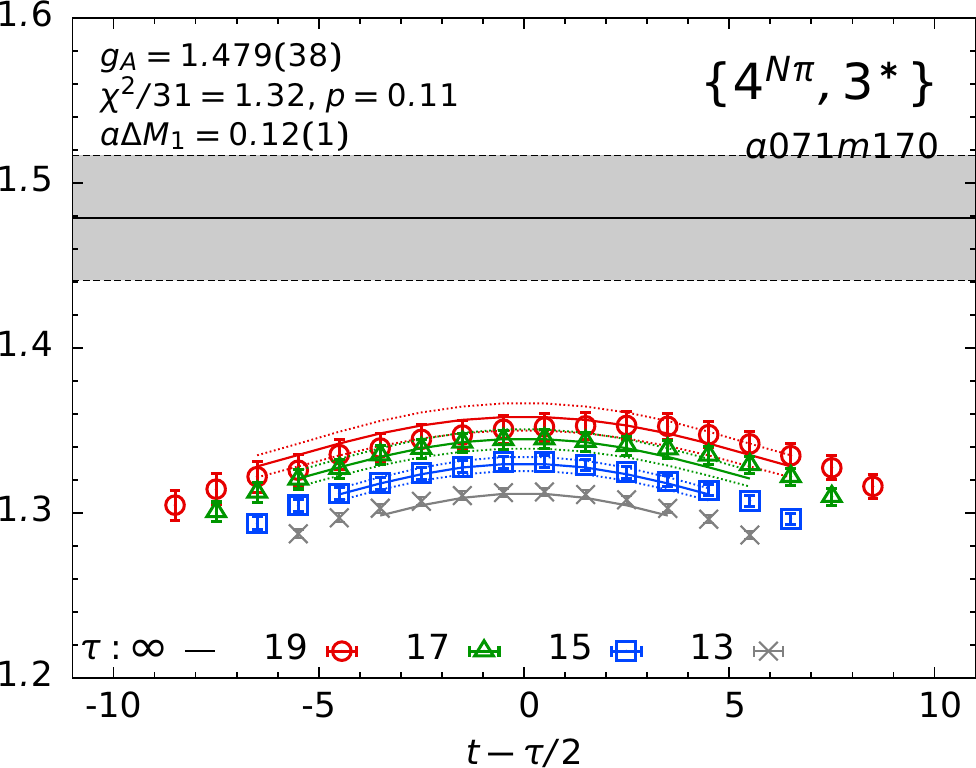}
\vspace{-0.1in}
\caption{Data for the ratio $\Gamma_A^3/\Gamma^2$ with the insertion
  of the axial current with $\vec p = 0$ at time $t$ are shown with different
  colors for 4 values of the source-sink separation $\tau$. Results for the ground state ($\tau \to \infty$) ME ($=g_A$),
  with two different plausible mass gaps, $a\Delta M_1$, in a 3-state truncation of 
  the spectral decomposition of $\Gamma_A^3$, are shown
  by the grey band, and for different values of $\tau$ by lines of the
  same color as the data. The two estimate of $g_A$ are different but the fits are not distinguished by  the $\chi^2/dof$. \looseness-1}
\label{fig:gAESC}
\end{figure}

The data in Fig.~\ref{fig:gAESC} for $g_A^{u-d}$ from a $\{a=0.071{\rm
  fm}, M_\pi = 170{\rm MeV}\}$ ensemble (see Ref.~\cite{Park:2021ypf}) illustrate
what the presence of ESC does  and our goal is to understand and reliably remove them. 
The data (same in the two panels) display the following
features of the ESC: 
\begin{itemize}
\item
The variation of the data with $t$ and $\tau$ is the signature of
ESC. In the limit $\tau-t$ and $\tau \to \infty$, the data
($\Gamma^3/\Gamma^2=g_A^{u-d}$) should be flat in $t$ and lie on top of each
other, i.e., independent of $\tau$ and $t$, particularly near $t
-\tau/2$, i.e., away from the source/sink.
\item
The data
should be symmetric about $t -\tau/2$ because $\Gamma^3$ is. The
statistical quality of the data for $\tau=19$ (=1.35 fm) is already
borderline in this respect.
\item
The convergence of the data with $\tau$ for fixed $t$ is monotonic and from below. This 
shows that ESC cause $g_A^{u-d}$ to be underestimated.
\item
ESC is removed by fitting data at the 3 largest values of
$\tau$ using the spectral decomposition given in
Eq.~\ref{eq:SF2and3pion} truncated at 3 states (ground plus two
excited states). The value for the ground state matrix element, given
by the fit, is shown by the grey band.\looseness-1
\item
The data in the two panels are the same. The fits differ in the energy, $E_1$, 
of the first excited state used. In the left panel it is the output of
the fit to $\Gamma^2$ while in the right, the energy of the lowest
$N\pi$ state with relative momentum $(0,0,1)$ is input using a narrow
prior.  The motivation for this including this $N\pi$ state is
$\chi$PT--- it contributes at one loop.
\item
The value of the ground state ME given by the two fits is different but the
augmented $\chi^2/dof$ are comparable. This highlights a serious problem: 
current data are not at sufficiently large $\tau$ nor precise
enough for fits to discriminate between different choices of $E_1$.
\item 
The two first excited-state energies, $E_1$, selected are, physics
wise, reasonable options: the first is given by the fit to $\Gamma^2$
and lies close to the N(1440), while in the second fit we input,
$N(0,0,1)\pi(0,0,-1)$. Furthermore, $N(0,1,1)\pi(0,-1,-1)$ and the rest
of the tower also contributes. In fact, all states with the same quantum
numbers contribute! What we do not know, a priori, are the amplitudes,
and the size of the contribution of each possible excited state to the ME.
In short, the statistical precision of the current data allow fits
with three states, however, these fits show that there are large regions in $E_1$
and $E_2$ that give similar $\chi^2/dof$ but significantly different
$\tau \to \infty$  values.
\end{itemize}

Bottom line: Until the data are good enough to distinguish between
fits with different number or combinations of plausible excited
states, and lacking a theoretical reason for a particular choice, the
difference between the extrapolated values with different possible
excited states can be regarded as an estimate of the systematic
uncertainty due to ESC.  A factor of 10 increase in statistics will give  data
for $\tau = \{21, 23\}$ with precision similar to $\tau = \{17, 19\}$ data shown in Fig.~\ref{fig:gAESC}. 
Then, I believe,  fits with  different $E_1$ will start to be discriminated by $\chi^2/dof$. 

Our data suggest that the uncertainty due to including or not the
$N(0,0,1)\pi(0,0,-1)$ state  could be a $\sim 5\%$ effect in
$g_A^{u-d}$, but is much smaller in $g_S^{u-d}$ or
$g_T^{u-d}$~\cite{Park:2021ypf}.  At this point in time, controlling
ESC remains the key outstanding systematic for achieving sub-percent
precision in the prediction of the isovector properties of nucleons.

\begin{figure}[h] 
    \includegraphics[width=0.32\linewidth]{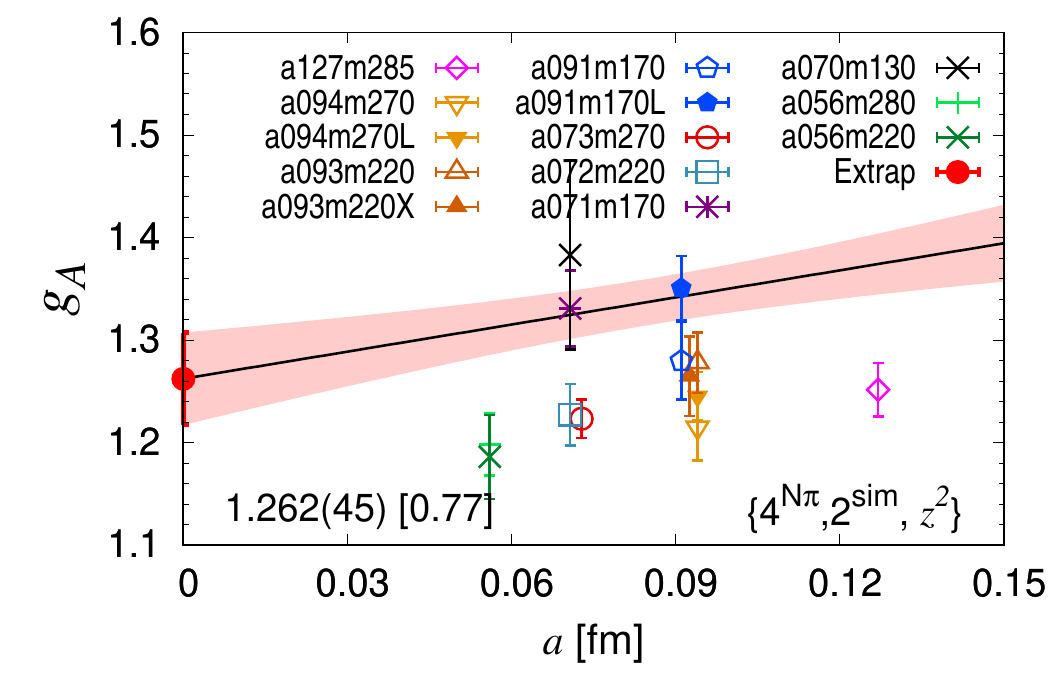}
    \includegraphics[width=0.32\linewidth]{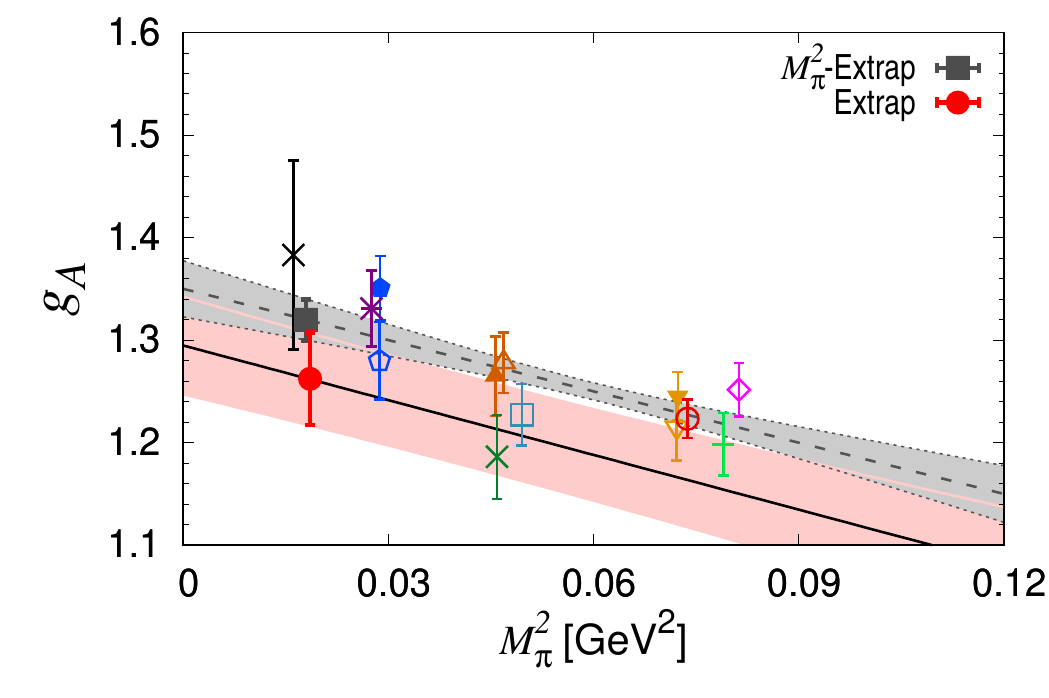}
    \includegraphics[width=0.32\linewidth]{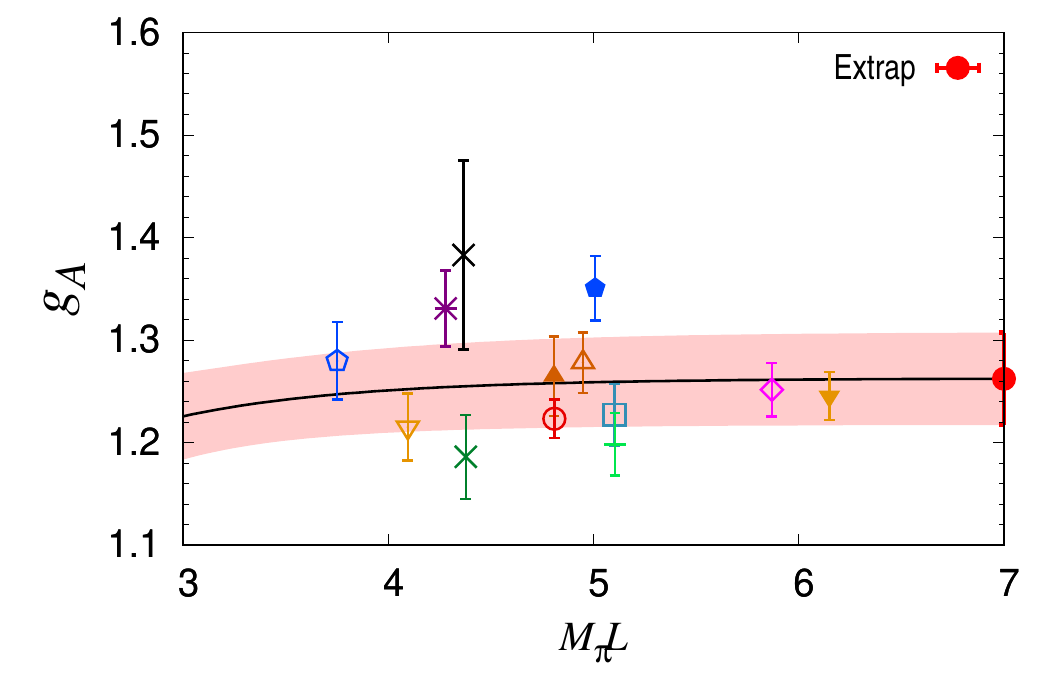}
\vspace{-0.1in}
\caption{Result of the simultaneous chiral-continuum-finite-volume
  fit, Eq.~\ref{eq:CCFV},  to $g_A^{u-d}$ data is shown by the pink band and
  plotted versus the lattice spacing $a$ (left), pion mass $M_\pi^2$
  (middle), and $M_\pi L$, the lattice size in units of $M_\pi$
  (right). The grey band in the middle panel shows a simple chiral fit, i.e., with 
  $c_2 = c_4 = 0$ in Eq.~\ref{eq:CCFV}. \looseness-1}
\label{fig:CCFV}
\end{figure}

The next step in the analysis, illustrated using the
$g_A^{u-d}$ data, is the chiral-continuum-finite-volume (CCFV) extrapolation
using a simultaneous fit in $\{M_\pi, a, M_\pi L\}$ shown in
Fig.~\ref{fig:CCFV}. The methodology is the same as described
in~\cite{Park:2021ypf}, except we now have data on 13 ensembles with
different $\{M_\pi, a, M_\pi L\}$. (Note, ${\overline m}_{ud}$ is specified
by $M_\pi$.) For the three iso-vector charges, the extrapolation ansatz, keeping lowest
order corrections in each of the 3 variables, is~\cite{Park:2021ypf}:
\begin{equation}
g(a,M_\pi,M_\pi L) = c_1+ c_2 a + c_3 M_\pi^2 +c_4\frac{M_\pi^2e^{-M_\pi L}}{\sqrt{M_\pi L}} \,.
\label{eq:CCFV}
\end{equation}
In such ans\"atz, the order of the correction terms one needs to include depends, in general, on the level of 
improvement of the lattice action and the observable.  
Each panel in Fig.~\ref{fig:CCFV} shows the result of this fit versus
a single variable with the other two set to their physical values. For
example,  in the panel versus $a$, the pink band shows the result with $M_\pi$ set to 
135~MeV and $M_\pi L \to \infty$. The data, however, have not been shifted 
in these two variables, which is why they do not lie within in the pink band.

Such simultaneous CCFV fits are routine  for getting physical values of observables.
Features in these fits that illustrate the size of the three systematics are
\begin{itemize}
\item
For the 2+1-flavor clover-Wilson action we have used, the value of
$g_A^{u-d}$ decreases by about 10\% as $a \to 0$, i.e., the slope with respect to $a$ is
positive.
\item
The value of $g_A^{u-d}$ increases by about 10\%  over the range of $M_\pi$ as $M_\pi \to 135$~MeV. 
\item
The dependence on $a$ and $M_\pi$ is largly independent of each other,
with opposite slopes.
\item
There are no significant finite volume corrections seen for $M_\pi
L > 4$.  This welcome feature has been observed in all calculations
involving single nucleon correlation functions.
\end{itemize}

Phenomenologically, the most interesting of the isovector charges is
the axial charge, $g_A^{u-d}$, which has been extracted from
experiments with high precision, $g_A^{u-d}/g_V =
1.2754(13)$~\cite{ParticleDataGroup:2022pth}. The precision and
robustness of lattice results have increased over the last decade, but
my conclusion, in light of possible unresolved ESC of multihadron
states such as the $N \pi$ that can cumulatively be as large as
$\approx 5\%$, is we need to better quantify and remove the ESC
contributions before we can claim sub-percent precision.

\vspace{-0.3in}

\section{Contribution of the Spin of the Quarks to the Nucleon Spin}
\label{sec:Nspin}

Using Ji's gauge invariant decomposition
${1}/{2} = \sum_{q=u,d,s,c} \left( {\Delta q}/2 + L_q \right) +
J_g $~\cite{Ji:1996ek}, where $L_q$ is the quark orbital and $J_g$ the gluon total
angular momentum, the contribution of the intrinsic spin of a quark with flavor $q$,
$\Delta q/2$, to the proton spin is given by a relation very similar
to Eq.~\ref{eq:gAdef}:\looseness-1
\begin{align}
\langle P(p,s') |  Z_A \, \bar{q} \gamma_\mu \gamma_5  q | P (p,s) \rangle =
g_A^q   \ \bar{u}_P (p,s') \, \gamma_\mu \gamma_5  \, u_P (p,s)  \,.
\label{eq:gAFDdef}
\end{align}
with $\Delta q = g_A^q$ and $u$ and $\bar u$ are the quark
spinors. Because the operator is diagonal in flavor, there is now an
additional Wick contraction in which the operator forms a closed loop
as illustrated in the right panel in Fig.~\ref{fig:2and3ptN}. This is
called a ``disconnected diagram''. The full contribution to $g_A^q$ is the sum of
the connected (middle) and disconnected (right) quark line diagrams.

The calculation of disconnected diagrams introduces a new layer of
computational cost.  The straightforward solution to calculate the
momentum projected quark loops with operator insertion on all time
slices is to calculate the all-to-all quark propagator $D^{-1}$. This 
is not practical as it is a  $(12 \cdot 10^{8}) \times (12 \cdot 10^{8}) $
complex matrix for a $100^4$ lattice.  The solution is to 
construct a stochastic estimate.  This approach works well, is
bias-free but introduces additional statistical uncertainty in the sum due to the
stochastic estimation of the disconnected quark loop diagram whereas the
calculation of the connected quark-line diagrams is exact up to matrix
inversion precision for $S_F$. This uncertainty in the measurement of the sum on
each configuration gets convoluted with that due to gauge fluctuations
in the ensemble average. Methods such as deflation and bias-corrected
truncated solver methods have allowed the reduction in errors in the
disconnected contributions to be of the same size as in the connected,
however, since their central value is smaller they contribute a larger
fraction to the overall error.

\begin{figure}[h]  
    \includegraphics[width=0.32\linewidth]{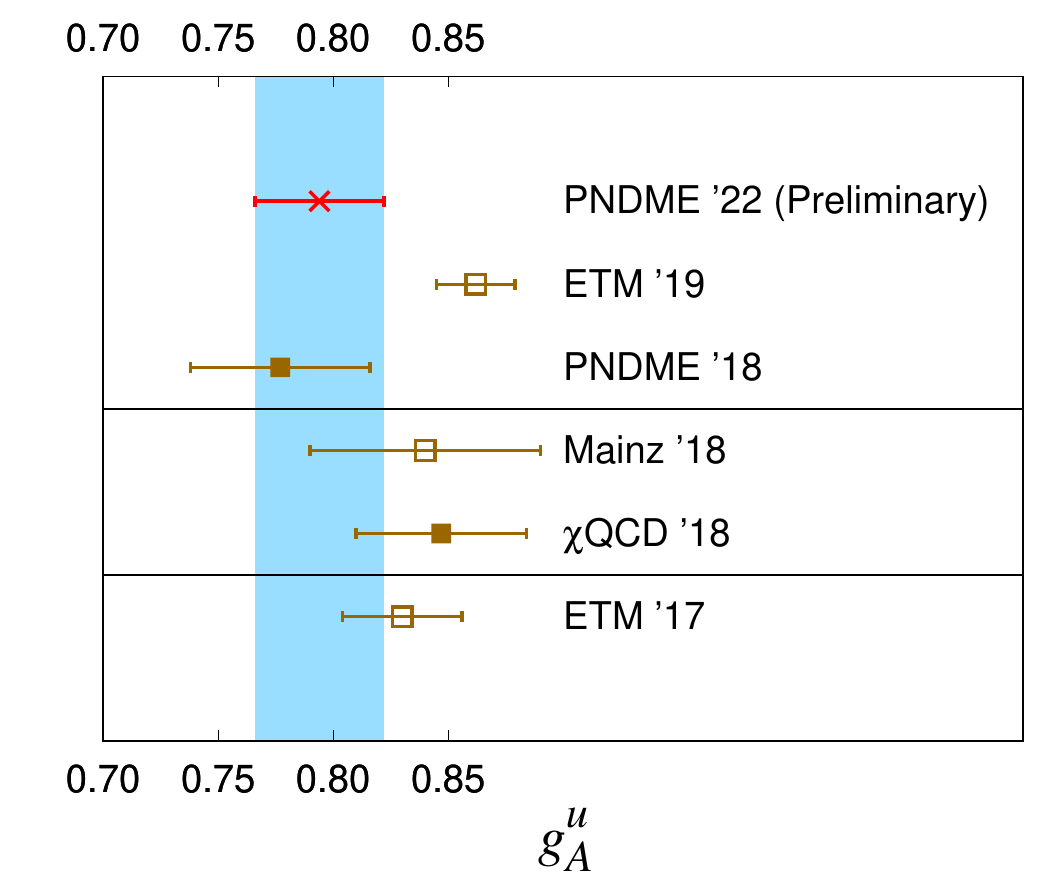}
    \includegraphics[width=0.32\linewidth]{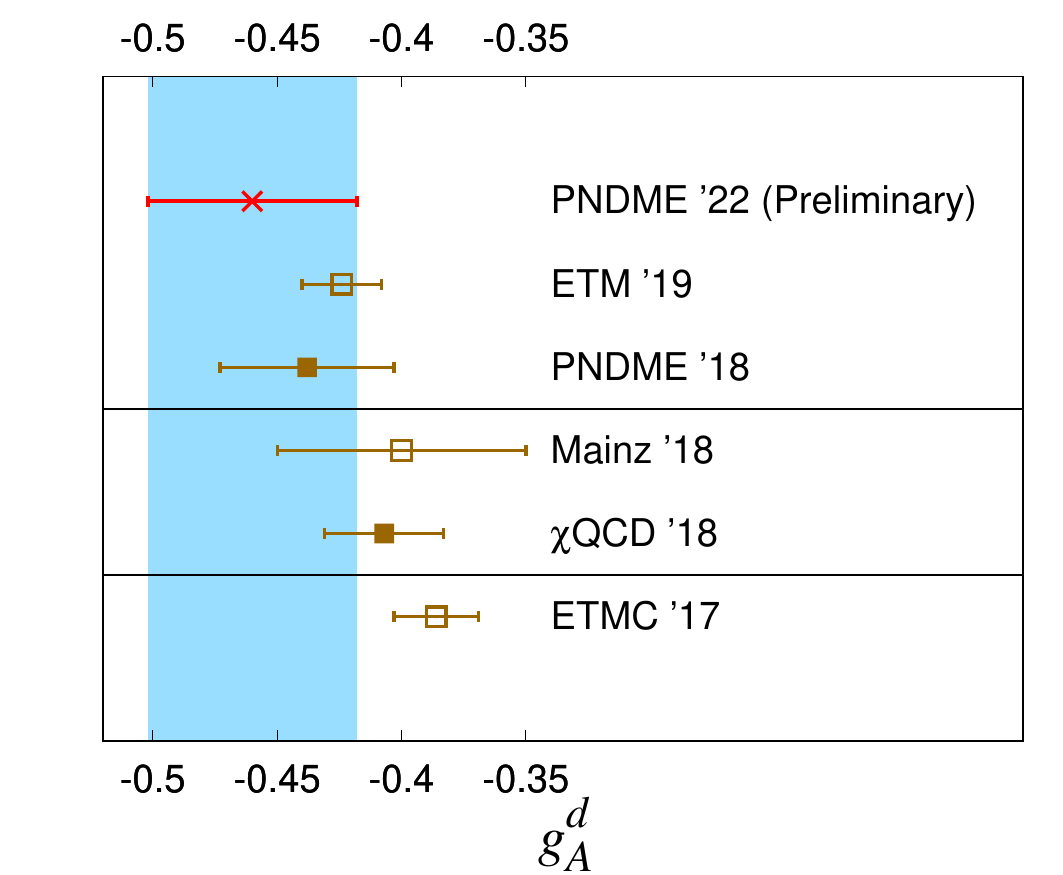}
    \includegraphics[width=0.32\linewidth]{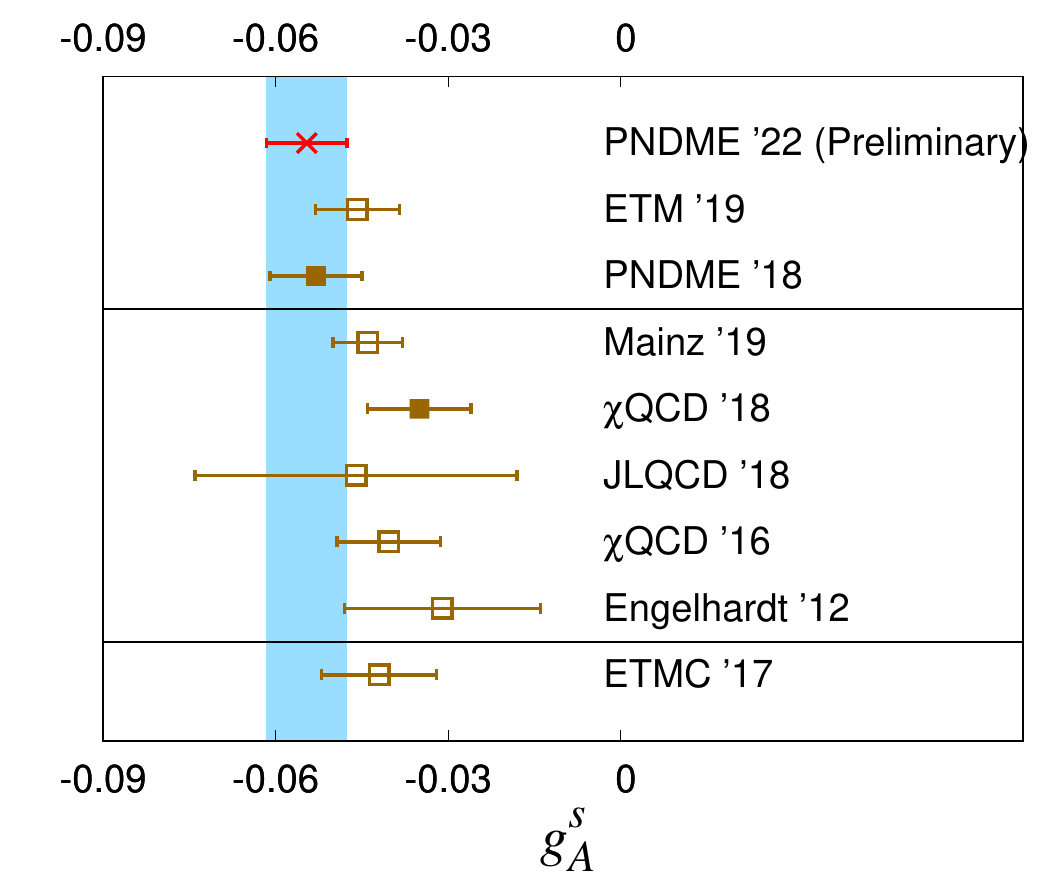}
\vspace{-0.1in}
\caption{Comparison of results for flavor diagonal axial charges of
  the nucleon in the FLAG
  format~\protect\cite{Aoki:2019cca,Aoki:2021kgd} obtained by
  different collaborations (
PNDME '22~\cite{Park:2021ggz}
ETM '19~\cite{Alexandrou:2019brg}, 
PNDME '18~\cite{Lin:2018obj}
Mainz '18~\cite{Djukanovic:2019gvi}, 
$\chi$QCD 18 \cite{Liang:2018pis}, 
JLQCD '18 \cite{Yamanaka:2018uud},
ETM '17 \cite{Alexandrou:2017oeh},
$\chi$QCD~15A~\cite{Gong:2015iir},
Engelhardt '12~\cite{Engelhardt:2012gd}). 
The points with filled squares meet the
  FLAG criteria for inclusion in the FLAG average. The contribution to the
  nucleon spin from quarks with flavor $q$ is $\Delta q /2 =
  g_A^q/2$. \looseness-1   }
\label{fig:gAfd}
\end{figure}

The steps in the analysis to get $g_A^u, \ g_A^d, \ g_A^s$ are the same as for the isovector charges described in
Sec.~\ref{sec:isovectorcharges}.  Our final results (PNDME 2018 and 2022 (preliminary))  
are shown in Fig.~\ref{fig:gAfd} along with those from other collaborations in the FLAG
format~\protect\cite{Aoki:2019cca,Aoki:2021kgd}.  In the FLAG review
process, results that pass the criteria for control over
discretization, finite lattice volume, renormalization and ESC systematic uncertainties,
and obtained sufficiently close to physical pion mass (or extrapolated
to $M_\pi = 135$~MeV), are then averaged and the overall error estimated with appropriate
consideration given to possible correlations between results. 
These FLAG
averages~\protect\cite{Aoki:2019cca,Aoki:2021kgd} are the community consensus value.

Our  (PNDME 2018 and 2022 (preliminary)) result  $\sum_q \Delta
q/2 = \sum_q g_A^q/2 = 0.14(3)$ is in good agreement with the extraction by the
COMPASS experiment $0.13< \Delta\Sigma/2 < 0.18$~\cite{Adolph:2015saz}.

\vspace{-0.3in}

\section{The Pion-nucleon Sigma Term}
\label{sec:piNsigma}

The pion--nucleon $\sigma$-term,   $\sigma_{\pi N}$, 
is a fundamental parameter of QCD that quantifies the amount of the
nucleon mass generated by up ($u$) and down ($d$) quarks having non-zero mass. In the "direct" method 
discussed here, it is, in the isospin symmetric limit
$\overline {m}_{ud} = (m_u + m_d)/2$,  given by\looseness-1
\begin{equation}
\sigma_{\pi N} \equiv {\overline   m}_{ud}\, g_S^{u+d} \equiv {m}_{ud} \allowbreak\, \langle N({\mathbf k},s)| \bar{u}
u + \bar{d} d | N({\mathbf k},s) \rangle \,,
\label{eq:sigma}
\end{equation}
where the extraction of ${\overline m}_{ud}$ and $g_S^{u+d}$ are done separately.
The scalar charge $g_S^q$ is determined
from the forward matrix element of the scalar density $\bar{q} q$
between the nucleon state:
\begin{align}
\bar{u}_N (0,s) g_S^q {u}_N (0,s) =  \langle N({\mathbf k}=0,s)| Z_S\ \bar{q}  q | N({\mathbf k}=0,s) \rangle,
\label{eq:gSdef}
\end{align}
where $Z_S$ is the renormalization constant and the nucleon spinor has
unit normalization. The scalar charges, $g_S^{q=u,d,s,c}$, also enters in the search for physics
beyond the Standard Model (SM): it determines the
coupling of the nucleon to any scalar mediator with quark content of
the coupling given by $\bar q q$, for example in direct-detection searches
for dark matter (DM) scattering off nuclei via a scalar mediator (similarly $g_A^q$ give the
spin dependent and $g_T^q$ the tensor couplings); in lepton
flavor violation in $\mu\to e$ conversion in nuclei; and in electric
dipole moments. The calculations of $g_S^q$ and $g_T^q$ are similar
to that discussed above for $g_A^q$. 

\begin{figure}[h]  
    \includegraphics[width=0.32\linewidth]{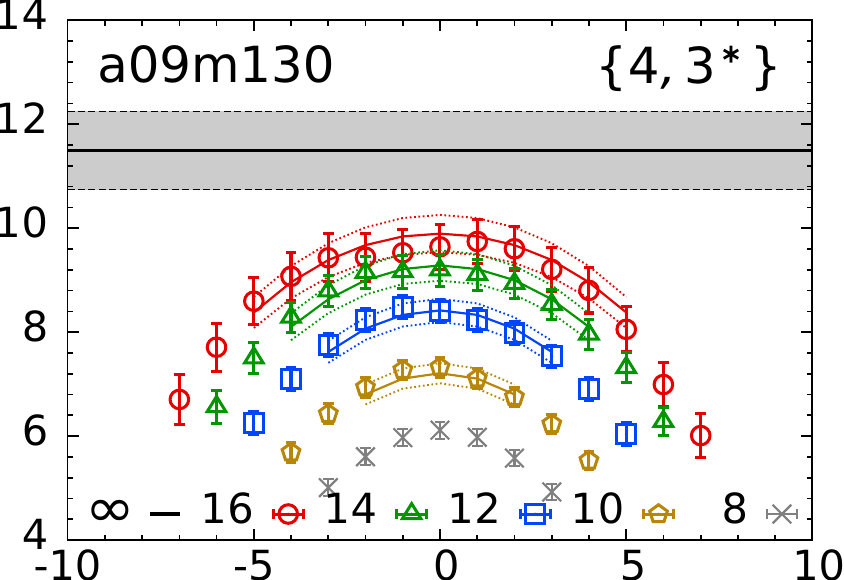}
    \includegraphics[width=0.32\linewidth]{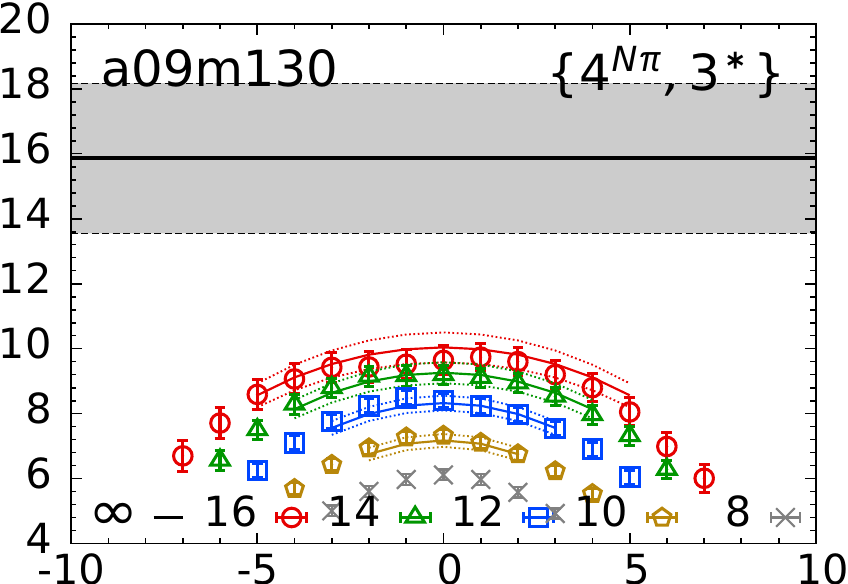}
    \includegraphics[width=0.32\linewidth]{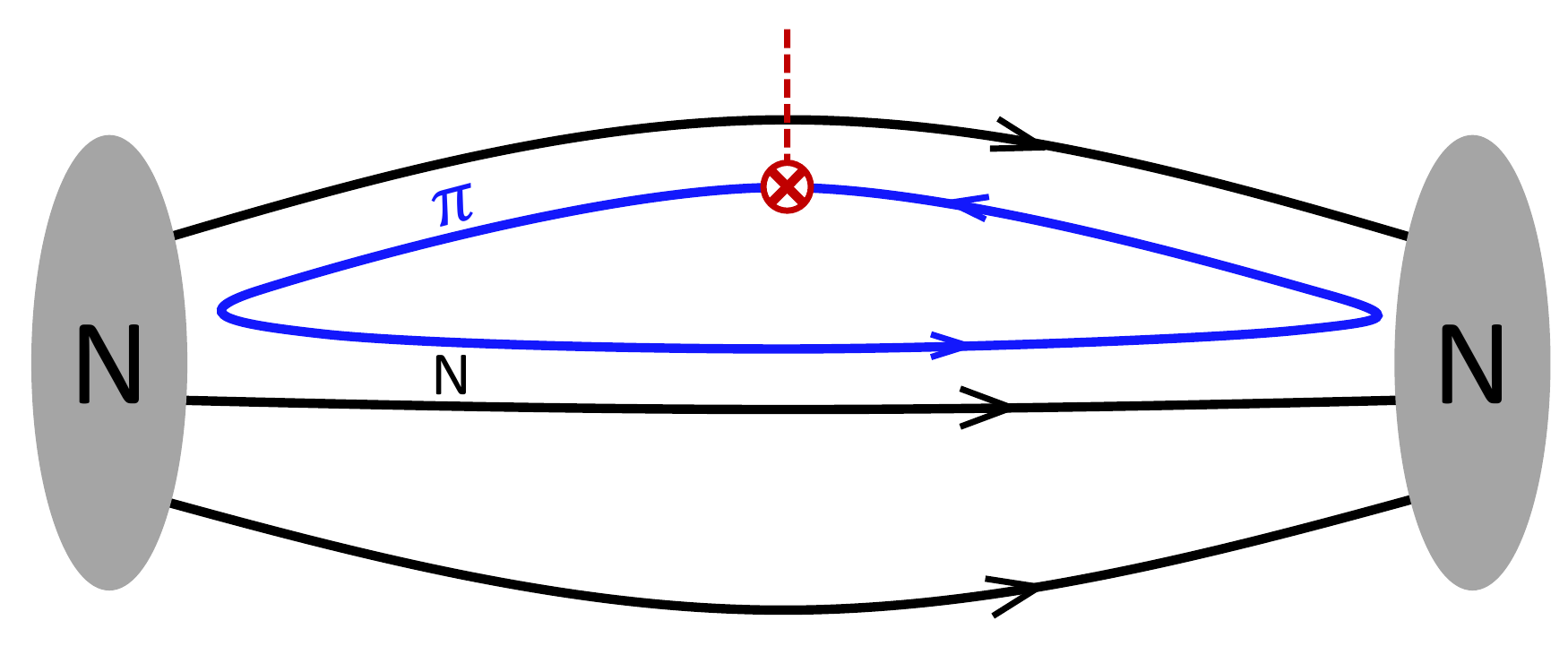}
\vspace{-0.1in}
\caption{Fits to get the scalar
  charges $g_S^u + g_S^d$ using for the mass gap $a \Delta M_1$ the value obtained from 
  $\Gamma^2$ (left) versus the noninteracting energy of the $N \pi$ state  (middle). 
  The right panel shows the disconnected diagram, 
  whose contribution is large.
  \looseness-1}
\label{fig:sigma}
\end{figure}

Figure~\ref{fig:sigma} shows data for $g_S^u + g_S^d$ from a
physical pion-mass ensemble~\cite{Gupta:2021ahb}.  Again the two fits
have very similar $\chi^2/dof$ but the one with $E_1$ of $N\pi$ as
the excited state gives almost 50\% larger value. Our NNLO $\chi PT$
analysis given in Ref.~\cite{Gupta:2021ahb} shows that there are two
significant ESC, one from $N\pi$ and the other from $N\pi \pi$, and each contribute 
about 10~MeV to $\sigma_{\pi N}$. These
two states are almost degenerate in our lattice calculation, so they
effectively contribute as one in the fit to $\Gamma^3$, i.e., their amplitudes in the spectral
decomposition add as the exponential factors are very
similar. The right panel in Fig.~\ref{fig:sigma} shows pictorially why
the $N \pi$ state makes a large disconnected contribution: the
scalar current has a large coupling to the quark loop, and the configuration shown in the
quark-line diagram favors a $N \pi$ intermediate state.

The two analyses with different values of $E_1$ led to an interesting
conundrum. The standard analysis (with $E_1 \sim  1450$~MeV) gave
$\sigma_{\pi N} \approx 40$~MeV consistent with previous lattice
analyses~\cite{Aoki:2019cca,Aoki:2021kgd}, whereas the analysis with
$E_1 =E_{N \pi} \approx 1230$~MeV gave $\sigma_{\pi N} \approx 60$~MeV, 
which is consistent with the dispersive analysis starting with the $N
\pi$ scattering data~\cite{Gupta:2021ahb}. Our preferred solution is the latter based on 
the $\chi PT$ analysis. In that case the tension between LQCD and 
phenomenological estimates is resolved. 

Clearly, the 50\% difference between the two analyses with similar
$\chi^2/dof$ calls for additional LQCD calculations to be done to
confirm this exciting result. The key point for future calculations of
$\sigma_{\pi N}$, using either the direct method of calculating the
charges $g_S^{u,d}$ as defined in Eq.~\ref{eq:sigma} or using the
Feynman-Hellmann relation, $\sigma_{\pi N} = m_q \partial M_N /\partial m_q$ 
where $M_N$ is the nucleon mass, is that they have to be done close to $M_\pi =
135$ as only there the $N \pi$ state becomes much lighter than
$N(1440)$ and the ESC are very different and manifest.  Our data with
$M_\pi > 200$~MeV do not give significantly different results between the two
kinds of fits. Thus, extrapolation from heavier $M_\pi$ ensembles will
miss this physics in both methods.\looseness-1

\vspace{-0.3in}

\section{Conclusions}

Simulations of LQCD provide ensembles of importance sampled
configurations whose distribution according to the Boltzmann factor
${\rm Det} {{\cal D}} e^{-A_{G}} = e^{-A_{G} + {\rm Ln} {\rm det}
  {\cal D}} $ constitutes the non-perturbative ground state of
QCD. This construction is exact (bias-free) but stochastic. For a
given lattice action, ensembles are characterized by six input
parameters $\{m_i, a, M_\pi L\}$ with $m_i \in \{m_u, m_d, m_s,
m_c\}$. Correlation functions of any time-ordered string of operators
are given by quark-line diagrams obtained using the Wick
contraction. Properties of QCD (spectrum, matrix elements, EoS, etc)
are extracted from expectation values, i.e., ensemble averages, of
these correlation functions.  The full excursion (possible ``paths'')
of the quark propagators (and values of link parameters for non-local
and gluon operators) in the quark line diagrams over 3-space but at a
fixed intermediate time generate the full Fock space wavefunction, a
linear combination of all states with the same quantum numbers as the
interpolating operator.  The propagation of each of these states in
Euclidean time $\tau$ is damped as $e^{-E_n \tau}$, allowing the
ground state with energy $E_0$ to be isolated in the limit $\tau \to
\infty$. Once again, this stochastic description of the wavefunction
provides no intuition or visualization but allows the calculation of
fully quantum mechanical matrix elements of any operator within this
state.\looseness-1

Lattice QCD is called a ``black box'' because we cannot
visualize or represent the vacuum fluctuations or the wavefuctions
that get created at intermediate times in the correlation functions
$\Gamma^n$. Nevertheless, the results obtained are rigorous, display
all the subtleties of QCD and confirm that this quantum field theory
describes the quantum dynamics  of quarks and gluons. 

Three kinds of observables have been used to
illustrate how LQCD calculations are done and data analyzed. Looking
ahead, precision calculations of nucleon correlation functions need to
overcome two challenges: the exponential degradation of the signal and
how to remove all/most excited state contamination in the wavefunctions to
get matrix elements within the nucleon ground state from correlations
functions calculated with finite source-sink separation $\tau$. The ESC
in current data can be large as illustrated by the pion-nucleon sigma term. 

The future is exciting -- the FLAG reports provide a growing testimony to 
LQCD having matured and results having an impact on phenomenology and
experiments~\cite{Aoki:2019cca,Aoki:2021kgd}.  Methodology and
algorithms for many calculations are robust, however, brute force
approach to achieving sub-percent precision in nucleon correlation functions and ME 
by just increasing the statistics is unlikely to succeed in the next few
years. It is, therefore, time for innovation and an exciting challenge to the
next generation--develop new methods and algorithms to reduce
systematics and increase statistics efficiently.

\vspace{-0.1in}
\section*{Acknowledgements and Funding Information}

Many thanks to my collaborators Tanmoy Bhattacharya, Vincenzo
Cirigliano, Martin Hoferichter, Yong-Chull Jang, Balint Joo, Huey-Wen Lin, Emanuele
Mereghetti, Santanu Mondal, Sungwoo Park, Frank Winter, Junsik Yoo,
and Boram Yoon with whom the work presented has been done over the
last decade.  The calculations used the CHROMA software suite. This
research used resources at (i) NERSC, a DOE Office of Science User
Facility supported by the Office of Science of the U.S. Department of
Energy under Award No. DE-AC02-05CH11231; (ii) the Oak Ridge
Leadership Computing Facility through ALCC award LGT107 and INCITE
award HEP133; (iii) the USQCD Collaboration, funded by DOE HEP; and
(iv) Institutional Computing at Los Alamos National Laboratory. R. Gupta is 
partially supported by DOE HEP under Award
No. DE-AC52-06NA25396 and by the LANL LDRD program. 

\vspace{-0.1in}

\bibliography{ref}

\nolinenumbers

\end{document}